\documentclass[twocolumn]{aastex63}

 \hypersetup{linkcolor=blue,citecolor=blue,filecolor=blue,urlcolor=blue}
\received{\today}
\revised{-- }
\submitjournal{AJ}

\shorttitle{Blazars overview with LAMOST}
\shortauthors{Pe\~na-Herazo et al.}
\newcommand{\fer}{{\it Fermi}}

\usepackage{graphicx}
\usepackage{longtable}
\usepackage{mdwlist}
\usepackage{natbib}
\usepackage{multirow}

\usepackage[switch]{lineno} % default option is 'left'
\usepackage{lineno}
\begin{document}

\title{An optical overview of blazars with LAMOST I: \\ Hunting changing-look blazars and new redshift estimates}

\correspondingauthor{Pe\~na-Herazo}
\email{harold.penaherazo@edu.unito.it }

\author[0000-0003-0032-9538]{Harold A. Pe\~na-Herazo}
\affiliation{Dipartimento di Fisica, Universit\`a degli Studi di Torino, via Pietro Giuria 1, I-10125 Torino, Italy.}
\affiliation{Instituto Nacional de Astrof\'{i}sica, \'Optica y Electr\'onica, Apartado Postal 51-216, 72000 Puebla, M\'exico. }

\author[0000-0002-1704-9850]{Francesco Massaro}
\affiliation{Dipartimento di Fisica, Universit\`a degli Studi di Torino, via Pietro Giuria 1, I-10125 Torino, Italy.}
\affiliation{Istituto Nazionale di Fisica Nucleare, Sezione di Torino, I-10125 Torino, Italy.}
\affiliation{INAF-Osservatorio Astrofisico di Torino, via Osservatorio 20, 10025 Pino Torinese, Italy.}
\affiliation{Consorzio Interuniversitario per la Fisica Spaziale (CIFS), via Pietro Giuria 1, I-10125, Torino, Italy.}

\author[0000-0002-4455-6946]{Minfeng Gu}
\affiliation{Key Laboratory for Research in Galaxies and Cosmology, Shanghai Astronomical Observatory, Chinese Academy of Sciences, 80 Nandan Road, Shanghai 200030, People's Republic of China    }

\author[0000-0002-5646-2410]{Alessandro Paggi}
\affiliation{INAF-Osservatorio Astrofisico di Torino, via Osservatorio 20, 10025 Pino Torinese, Italy.}
\affiliation{Consorzio Interuniversitario per la Fisica Spaziale (CIFS), via Pietro Giuria 1, I-10125, Torino, Italy.}

\author[0000-0003-2204-8112]{Marco Landoni}
\affiliation{INAF-Osservatorio Astronomico di Cagliari, via della Scienza , 5, Selargius, CA, Italy.}
\affiliation{INAF-Osservatorio Astronomico di Brera, Via Emilio Bianchi 46, I-23807 Merate, Italy.}

\author[0000-0003-3073-0605]{Raffaele D'Abrusco}
\affiliation{Center for Astrophysics | Harvard \& Smithsonian, 60 Garden Street, Cambridge, MA 02138, USA.}

\author[0000-0001-5742-5980]{Federica Ricci}
\affiliation{Dipartimento di Fisica e Astronomia, Università di Bologna, via P. Gobetti 93/2, I-40129, Bologna, Italy.}  
\affiliation{INAF-Osservatorio di Astrofisica e Scienza dello Spazio, via Gobetti 93/3, I-40129, Bologna, Italy.}

\author[0000-0001-9487-7740]{Nicola Masetti}
\affiliation{INAF-Osservatorio di Astrofisica e Scienza dello Spazio, via Gobetti 93/3, I-40129, Bologna, Italy.}
\affiliation{Departamento de Ciencias F\'isicas, Universidad Andr\'es Bello, Fern\'andez Concha 700, Las Condes, Santiago, Chile. }

\author[0000-0002-2558-0967]{Vahram Chavushyan}
\affiliation{Instituto Nacional de Astrof\'{i}sica, \'Optica y Electr\'onica, Apartado Postal 51-216, 72000 Puebla, M\'exico. }
          



\begin{abstract}

The extragalactic $\gamma$-rays sky observed by \fer-Large Area Telescope (LAT) is dominated by blazars. In the fourth release of the \fer-LAT Point Source Catalog (4FGL), are sources showing a multifrequency behavior similar to that of blazars but lacking an optical spectroscopic confirmation of their nature known as Blazar Candidate of Uncertain type (BCUs).
We aim at confirming the blazar nature of BCUs and test if new optical spectroscopic observations can reveal spectral features, allowing us to get a redshift estimate for known BL Lac objects. We also aim to search for and discover changing-look blazars (i.e., blazars that show a different classification at different epochs).
We carried out an extensive search for optical spectra available in the Large Sky Area Multi-Object Fibre Spectroscopic Telescope (LAMOST) Data Release 5 (DR5) archive.  
We selected sources out of the 4FGL catalog, the list of targets from our follow-up spectroscopic campaign of unidentified or unassociated $\gamma$-ray sources, and the multifrequency catalog of blazars: the Roma-BZCAT. We selected a total of 392 spectra. We also compare some of the LAMOST spectra with those available in the literature.
We classified 20 BCUs confirming their blazar-like nature. Then we obtained 15 new redshift estimates for known blazars. We discovered 26 transitional (i.e., changing-look) blazars that changed their classification. Finally, we were able to confirm the blazar-like nature of six BL Lac candidates. All remaining sources analyzed agree with previous classifications.
BL Lac objects are certainly the most elusive type of blazars in the $\gamma$-ray extragalactic sky. 
\end{abstract}



\vspace{3 mm}

\section{Introduction} 
Powered by a supermassive black hole lying in the center of elliptical galaxies, blazars are a subclass of radio-loud active galaxies mainly characterized by non-thermal emission that dominates their broad-band spectral energy distribution, although blazars can be dominated by thermal emission in optical to ultraviolet frequencies \citep{giommi95,fossati98,abdo10,1lac,mao16}. They are certainly the rarest class of extragalactic radio sources \citep{urry95} but, on the other hand, are also the largest known population of gamma-ray objects \citep{3fgl,4fgl,massaro15}. 

Blazars feature peculiar, multifrequency properties as (i) flat radio spectra both below \citep{massaro13a,nori14,giroletti16} and above $\sim$1 GHz \citep{healey07,petrov13,schinzel15,schinzel17}, (ii) radio to optical polarization \citep{poutanen94,park18,mandarakas19,liodakis19}, (iii) super-luminal motion seen at radio frequencies \citep{jorstad01,kellermann07,lister13}, (iv) infrared colors, not simply ascribable to dust emission, \citep{massaro11,dabrusco12}, (v) bolometric luminosities up to 10$^{46-48}$ erg/s \citep{zhang12} and (vi) both intensity and spectral variability at all frequencies from radio and optical \citep{gu11,chatterjee12,leon13,isler13,hayashida15,patino17,sarkar19,nalewajko19,zhang20,chavushyan20} up to TeV energies \citep{giannios09,acciari10,archambault15} and with daily to minutes timescales \citep{aharonian07,albert07,paliya17,liu17,gupta18}, all coupled with a typical double-humped broadband spectral energy distribution (SED) \citep{abdo10,fan16}. In 1978, at the Pittsburg conference, when the word {\it blazar} was coined, Blandford and Rees proposed the current interpretation of blazar observational properties due to relativistic particles accelerated in a plasma jet closely aligned to the line of sight \citep{blandford78}.

From an optical perspective, blazars are mainly divided into two sub-classes: BL Lac objects and flat-spectrum radio quasars (FSRQs). The former sub-class shows an almost featureless spectrum with a relatively blue continuum where only weak emission or absorption lines are, rarely, superimposed and, when occurs, with an equivalent width less than $\sim$5 \AA\ \citep{stickel91,stocke91}. On the other hand, the latter sub-class present the typical quasar-like spectra with relatively intense and broad emission lines and the presence of the big blue bump peaking at ultraviolet wavelengths \citep[see, e.g.,][]{wu12,shaw12}. 
According to the multifrequency catalog of blazars, the Roma-BZCAT, the former sub-class, is generally indicated as BZB, while the latter as BZQ \citep{bzcat1}. Then in the recent release of the Roma-BZCAT \citep{bzcat5}, radio sources, usually reported as BL Lac objects in the literature, but showing a SED where the host galaxy emission is significantly dominant over the continuum due to the relativistic jet, are indeed labeled as BZG.   
The Ca II break is usually adopted to evaluate the contribution of non-thermal continuum with respect to the host emission, and thus differentiate between BZGs and BZBs \citep{landt02,massaro14a}. 
The Ca II break is defined as $C = (F_+-F_-)/F_+$, where $F_+$ and $F_-$ are the fluxes at rest-frame wavelengths of 3750-3950 \AA\, and 4050-4250 \AA. Blazars with a $C\geq0.25$ are classified as BZG while those with $C<0.25$ as BZB \citep{bzcat5}.

The spectral variability of blazars provides crucial information to study the nature of their broad line region  (BLR). \citet{leon13} reported the statistically significant flare-like event in the Mg {\sc II} emission line in the blazar 3C 454.3 and  presented direct observational evidence of the BLR close to the radio core of the jet, and this was confirmed in subsequent studies \citep{isler13,jorstad13,amaya21}. Recently, similar behavior was observed for CTA 102 but on a greater scale \citep{chavushyan20}. These blazars are the only ones where an increase in the emission line (Mg {\sc II} and Fe {\sc II} band) was reported. Both blazars, 3C 454.3 and CTA 102 are FSRQ type.

The last decade has undoubtedly been a golden age for blazar research, due to the all-sky survey of the \fer\ satellite in the MeV-GeV energy range, carried out thanks to its Large Area Telescope \citep[LAT;][]{atwood09}. This allowed us to discover hundreds of new blazars associated with previously unidentified/unassociated $\gamma$-ray sources  \citep[UGSs; see][for reviews on the extragalactic sky seen by \fer]{massaro15, pena20}. Since the release of the First \fer-LAT Point Source Catalog \citep[1FGL;][]{1fgl} and until the latest, the Fourth \fer-LAT Point Source catalog \citep[4FGL;][]{4fgl} it was quite clear that an almost constant fraction, about 1/3, of all detected objects, were and still are UGSs \citep{quest}, lacking an assigned low energy counterpart \citep{pena20}, or being simply unclassified mainly due to the lack of an optical spectrum, recently labeled as Blazar of Uncertain type \citep[BCUs; see also][]{3fgl, 3lac}. Hundreds of blazars were discovered thanks to new follow-up observations of UGSs and BCUs available at radio \citep{kovalev09,petrov13,massaro13a,nori14,schinzel15,giroletti16}, optical \citep{paiano17b} and X-ray \citep{cheung12,paggi13,acero13,takeuchi13,landi15,kaur18,kaur19,marchesini19a,marchesini20a} frequencies as well as applying statistical methods \citep{ackermann12a,doert14,salvetti17}, hundreds of them are probably still unknown \citep{massaro12a,dabrusco13}.

In between the releases of the Second and the Third \fer\ Point Source catalogs \citep[2FGL and 3FGL, respectively][]{2fgl,3fgl}, thanks to the discovery that blazars have extremely peculiar mid-IR colors \citep{massaro11,dabrusco12,massaro12b,massaro16a,dabrusco19}, we were able to find hundreds of potential blazar-like counterparts of UGSs \citep{quest,pena20}. Then thanks to an extensive follow up spectroscopic campaign in the optical band \citep{optcmpi,optcmpii,optcmpiii,optcmpiv,optcmpv,optcmpvi,optcmpvii,optcmpviii,optcmpix,optcmpx} discovering and classifying hundreds of blazars as summarized in \citet{pena20}. We classified more than 400 new blazars, mainly belonging to the BL Lac subclass, the most elusive one. The results of our campaign also included a search in the archives of optical surveys \citep{crespo16} as the Sloan Digital Sky Survey \citep[SDSS][]{aguado19} and the Six-Degree Field Galaxy Survey \citep[6dFGS;][]{jones04,jones09}. 

%general:
%in the intro at some point you mght add a sentence or two about the impact of the spectroscopic analysis of blazars and cite papers for the LF, background gamma, TeV candidates, radio weak BL Lacs etc. be sure you add these two:
Clarifying the nature of blazar candidates and determining the redshift of the $\gamma$ blazar population is crucial to (i) determine their luminosity function \citep{ajello14,ackermann16}, (ii) study the imprint of the extragalactic background light in the blazar $\gamma$-ray spectra \citep[e.g.,][]{dominguez11,ackermann12b,sandrinelli13} (iii) select potential targets for TeV observatories \citep{massaro13d,arsioli15} (iv) search for new classes of $\gamma$-rays sources \citep{massaro17,bruni18} (v) analyze new methods for $\gamma$-rays detection \citep{4fgl,kerr19}, and (vi) set constraints on the annihilation of dark matter in subhalos \citep[see e.g.,][]{zechlin12,ackermann14,berlin14}.

Here we propose to analyze the optical spectra of several blazars samples using archival data collected by the Large Sky Area Multi-Object Fibre Spectroscopic Telescope \citep[LAMOST,][]{lamost0,lamost1,lamost2}. Our goals are to confirm the blazar nature of BCUs, to get a redshift estimate for known BL Lac objects if new optical observations can reveal the presence of spectroscopic features, and search for changing-look blazars.

We aim at exploring not only $\gamma$-ray sources classified in the 4FGL as BCUs, for which the assigned counterpart lacks an optical spectroscopic classification but also to verify the classification of all $\gamma$-ray blazars observed during our follow-up campaign and those found in the literature \citep[see e.g.,][]{shaw13,paiano17b,klindt17,desai19} and blazars listed in the Roma-BZCAT, all lying in the LAMOST footprint. Moreover, we could detect any spectral variability between the epoch when the optical spectra were collected.

The LAMOST telescope (also known as the Guoshoujing telescope, GSJT) is a reflective Schmidt telescope with active optics with a total $5\times5$ deg$^{2}$ field of view. On the focal plane, there are 4000 fiber-positioning units with a size of 3.3 arcseconds. Each unit feeds an optical fiber which transfers light to one of sixteen 250-channel spectrographs \citep[see][for additional details and configuration]{lamost0,lamost1,lamost2}. The spectrographs have two resolution modes low resolution with R $\sim$ 1000 and wavelength coverage of 3700 to 9100 \AA\, and medium resolution mode with R$\sim$5000.

Archival spectra collected here are part of the wide-field survey, called the LAMOST Experiment for LAMOST ExtraGAlactic Surveys (LEGAS), including as scientific objective an extra-galactic spectroscopic survey to shed light on the large scale structure of the universe \citep{lamost2}.  LEGAS plans to acquire spectra of galaxies with magnitudes up to $r=19.5$ with a sky coverage of 11500 deg$^2$, with declination $-10 < \delta < 60$. We used the LAMOST data release (DR) 5 \citep{lamostdr5} that includes 152863 galaxies and 52453 quasars.

This paper is organized as follows: in Section~\ref{sec:sample} we describe our sample selection. Then, in Section~\ref{sec:results}, we present results achieved thanks to the spectral analysis of archival LAMOST spectra. Finally, Section~\ref{sec:summary} is devoted to our summary, conclusions, and future perspectives.

%We use cgs units unless stated otherwise. Spectral indices, $\alpha$, are defined by flux density, S$_{\nu}\propto\nu^{-\alpha}$.

\section{Sample Selection}
\label{sec:sample}
We extracted sources to carry out our analysis from the following three samples or catalogs. 

\begin{enumerate}
\item The first catalog is the 4FGL \citep{4fgl} based on the first eight years of science data collected by Fermi in the energy range between 50 MeV to 1 TeV. The 4FGL catalog includes 5064 sources above four sigma significance, including 75 sources that are spatially extended, 358 considered identified based on the angular extent, periodicity, or correlated variability observed at other wavelengths, and 1336 lacking plausible counterparts at shorter wavelengths. The 4FGL includes more than $\sim$3100 sources, either identified or associated, with known active galaxies belonging to the blazar class and 239 pulsars. 
\item We selected the 517 sources listed in our optical follow-up campaign to unveil the nature of UGSs and BCUs \citep{cowperthwaite13,optcmpi,optcmpii,optcmpiii,quest,optcmpiv,optcmpv,optcmpvi,optcmpvii,optcmpviii,optcmpix,optcmpx,pena20}. This campaign, carried out during the last decade, significantly augmented the number of known sources in the 4FGL, discovering and classifying 394 targets \citep[see also,][]{massaro13b,massaro15b,massaro16b,demenezes19} with an additional 123 sources with spectroscopic information collected from a literature search \citep[e.g.,][]{shaw13,klindt17,paiano17b,paiano17c,paiano17d,desai19}. 
\item Finally, we also verify redshifts and classification of those blazars listed in the Roma-BZCAT v5.0 \citep{bzcat5}, the latest release, including a total of 3561 sources, divided as 1151 BZBs, 369 with a firm redshift estimate, 1909 BZQs, and 274 BZGs.
\end{enumerate}   

However, not all these sources were observed spectroscopically in the optical band, and thus not all of them have an available spectrum in the LAMOST DR 5 used in our analysis. Thus restricting to the LAMOST footprint, we extracted the following samples out of each catalog previously indicated.
\begin{enumerate}
\item In the 4FGL, we selected a total of 31 sources classified as one AGN, 19 BCUs, six BL Lacs, and five flat-spectrum radio quasars, the latter two classes labeled as BLL and FSRQ therein.
\item From the 517 blazars observed during the optical spectroscopic campaign of UGSs and BCUs, only eight sources have good quality spectra available in the LAMOST footprint.
\item Then, in the Roma-BZCAT, we found a total of 353 blazars investigated. These were classified as 130 BZBs, 184 BZQs, and 39 BZGs.
\end{enumerate} 
%It is worth noting that when we found one source that is listed in both the 4FGL or observed during our spectroscopic campaign but included already in the Roma-BZCAT, we kept it only in the latter sample.
In those cases of sources overlapping within our three initial samples, we only indicated and list them as part of the Roma-BZCAT.

All numbers reported above did not include LAMOST spectra available with a low signal to noise ratio (i.e., below ten measured between 4000 \AA\ and 9000 \AA) since they do not allow us to obtain a precise classification. It is worth noting that these numbers do not include repeated multiple matches in the considered samples, and a maximum angular separation of 2 arcsec was used to search for a LAMOST counterpart. We determined the choice of 2 arcsec as angular separation to associate each source to its LAMOST counterpart adopting the same statistical procedure described in \citep{massaro14b} for the blazar counterpart in the SDSS and here corresponding to a chance probability of spurious associations lower than $\sim$1\%. 

We list all the sources from the three samples in Table~\ref{tab:results}, along with all our results. Here we report (1) the sample each source belong to; (2) source name, (3) class and (4) redshift in the original catalog; (5) the LAMOST designated name; (6) the class and (7) the redshift identified thanks to our analysis; and, (8) a flag indicating if the source also has an available spectrum in the SDSS that we used for comparison.

\section{Analysis, Classification and Redshift estimates}
\label{sec:results}
For all sources in our three samples, we retrieved the optical spectrum available in the LAMOST DR 5 and performed the spectral analysis to measure emission or absorption lines eventually detectable. 
We searched for Balmer emission lines or other characteristic spectral features, as C {\sc IV}, C {\sc III}], Mg {\sc II}, [O {\sc II}], H$\beta$, [O {\sc III}], H$\alpha$ and [N {\sc II}] complex, or the [S {\sc II}] doublet in emission, Ca {\sc II} H\&K, the G band, or Mg {\sc II} doublet in absorption.

Data analysis was then carried out using IRAF standard packages \citep{tody86}. We measured the equivalent width at the observer frame, defining a local continuum. We measured redshift for those sources we identified at least three absorption or emission lines. The line positions were estimated fitting a Gaussian profile near the peak of the line to avoid possible shifted spectral components.
We estimated lower limits on the redshifts for those BZBs having detection of intervening absorption systems, as previously seen in BL Lac spectra \citep{stocke97,paiano17c,landoni18,landoni20,paiano20}.

\begin{figure*}
\begin{center}
\includegraphics[width=\textwidth]{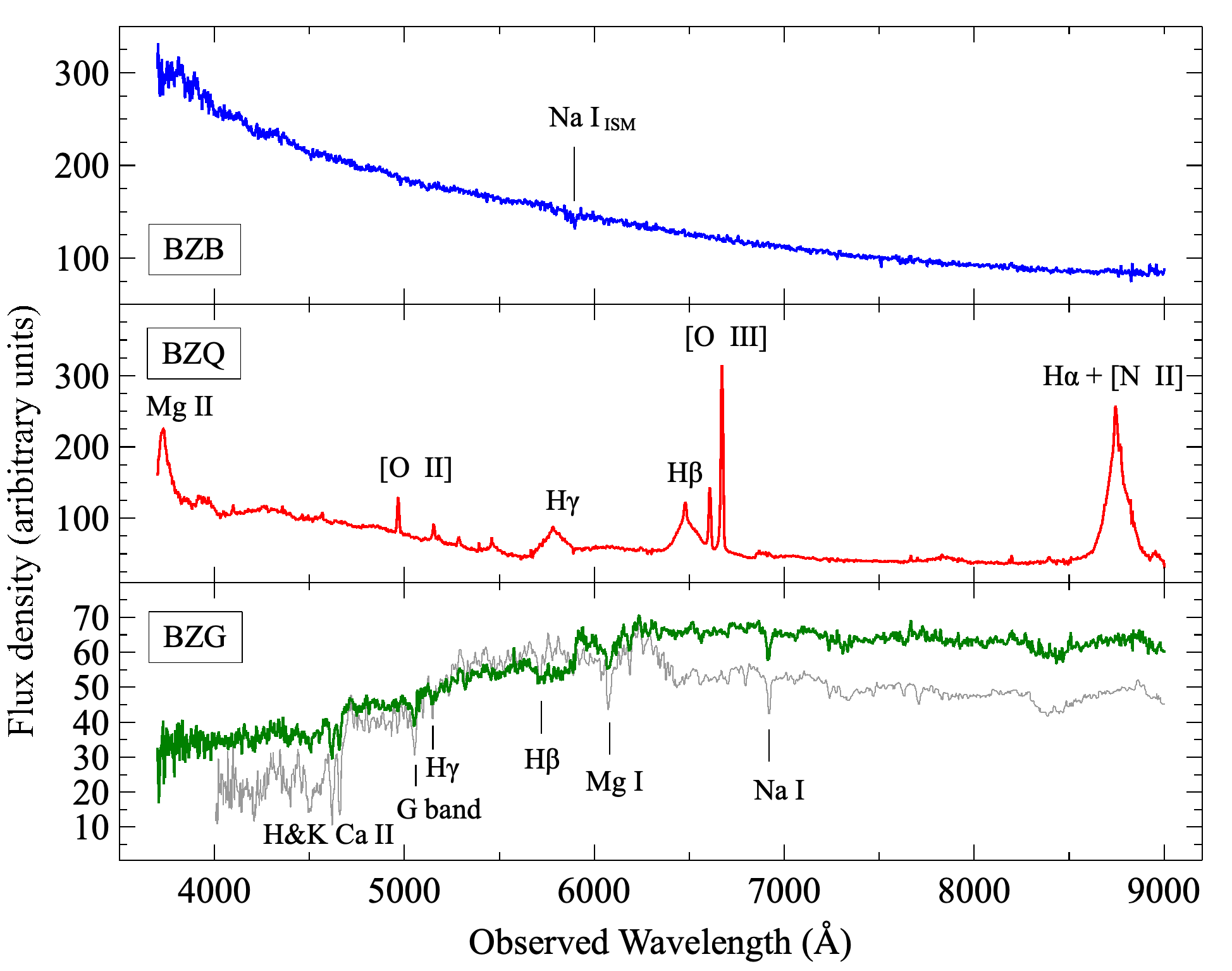}
\end{center}
\caption{(Upper panel) LAMOST spectrum of J065046.49+250259.6, a classical BZB, showing the almost featureless blue spectrum, the only absorption feature is the galactic interstellar absorption of Na I. 
(Central panel) LAMOST spectrum of J011218.05+381856.8 a source classified as BZQ, showing a quasar-like spectrum along with labels of some emission lines. 
(Lower panel) In green, the spectrum of a BZG, i.e., J090536.44+470546.3 analyzed during our investigation. In grey, we compare the BZG with the spectrum of an elliptical galaxy, J084640.34+281829.6, rescaled to match the flux of the BZG at 4730 \AA.
%All these spectra are provided to illustrate as templates of those analyzed here. 
}
\label{fig:exbzqbzb}
\end{figure*}

Since we are searching for blazars, we adopted the following classification, based on the criteria of the Roma-BZCAT. We classified and labeled BL Lac objects as BZBs (featureless spectra or with emission lines of EW $<$ 5 \AA), flat-spectrum radio quasars as BZQs (quasar-like spectra), and indicated as BZGs those blazars having a non-negligible host galaxy emission in both their optical spectra and in their broadband SED \citep{massaro14a}. 
BZBs have almost featureless optical spectra with only weak absorption or emission lines of equivalent width less than 5 \AA\ \citep{stickel91,stocke91}, BZQs show typical quasar-like optical spectra while BZGs are more similar to classical elliptical galaxies in the optical band \citep{shaw12,massaro14a,bzcat5}, same classification scheme was also used in our previous analyses \citep{pena20,optcmpx,quest}. In Figure~\ref{fig:exbzqbzb} we show all spectra of sources belonging to each blazar class used in our analysis: a BL Lac object, a quasar-like spectrum typical of BZQs, the one of a classical elliptical galaxy as those of BZGs. 
%Then in Table~\ref{tab:results}, we present all our results. Here we report (1) the sample each source belong to; (2) source name, (3) class and (4) redshift in the original catalog; (5) the LAMOST designated name; (6) the class and (7) the redshift identified thanks to our analysis; and, (8) a flag indicating if the source also has an available spectrum in the SDSS that we used for comparison.
%Then in Table~\ref{tab:summary}, we summarize our classification results, grouping the sources by their original sample. We describe our results in the following.
In the next subsections, we describe the LAMOST spectral analysis results separately for our three catalogs (also summarized in Table~\ref{tab:summary}).

%------------------------------------------------------------------------------------------------------------------------
\begin{table*}
%\label{tab:log}
\caption{Classification results}
\resizebox{\textwidth}{!}{
\begin{tabular}{llcllclc}
\noalign{\smallskip}
\hline
\noalign{\smallskip}
\hline
  \multicolumn{1}{c}{Sample} &
  \multicolumn{1}{c}{Name} &
  \multicolumn{1}{c}{Cat. Class} &
  \multicolumn{1}{c}{z$_1$} &
  \multicolumn{1}{c}{Designation} &
  \multicolumn{1}{c}{Our Class} &
  \multicolumn{1}{c}{z$_2$} &
  \multicolumn{1}{c}{SDSS flag} \\

  \multicolumn{1}{c}{(1)} &
  \multicolumn{1}{c}{(2)} &
  \multicolumn{1}{c}{(3)} &
  \multicolumn{1}{c}{(4)} &
  \multicolumn{1}{c}{(5)} &
  \multicolumn{1}{c}{(6)} &
  \multicolumn{1}{c}{(7)} &
  \multicolumn{1}{c}{(8)} \\
\hline
BZCAT & 5BZQ J0005+0524 & bzq & 1.900 & J000520.21+052410.7 & bzq & 1.900 & \dots\\
BZCAT & 5BZQ J0005+3820 & bzq & 0.229 & J000557.17+382015.1 & bzq & 0.230 & \dots\\
BZCAT & 5BZG J0006+1051 & bzg & 0.168 & J000620.30+105151.1 & bzb & 0.168 & \dots\\
BZCAT & 5BZQ J0010+1724 & bzq & 1.601 & J001033.99+172418.7 & bzq & 0.769 & \dots\\
BZCAT & 5BZB J0015+3536 & bzb & \dots & J001527.94+353639.0 & bzb & \dots & \dots\\
BZCAT & 5BZQ J0019+2602 & bzq & 0.284 & J001939.75+260252.8 & bzq & 0.284 & \dots\\
4FGL & 4FGL J0021.0+0322 & bcu & \dots & J002050.26+032358.2 & bzb & \dots & \dots\\
BZCAT & 5BZG J0022+0006 & bzg & 0.306 & J002200.95+000657.9 & bzb & 0.306 & \dots\\
4FGL & 4FGL J0028.9+3553 & bcu & \dots & J002851.97+355036.0 & bzb & \dots & \dots\\
BZCAT & 5BZQ J0029+0554 & bzq & 1.314 & J002945.88+055440.6 & bzq & 1.314 & \dots\\
\hline
\hline
\noalign{\smallskip}
\end{tabular}}
\label{tab:results}
Note: Columns (1) Original sample; (2) Source name ; (3) class and (4) redshift in the sample catalog; (5) the LAMOST designated name; (6) class and (7) the red-shift identified thanks to our analysis; and (8) flag indicating if the source has also an available spectrum in the SDSS.
This table is available in its entirety in machine-readable form.
\end{table*}
%------------------------------------------------------------------------------------------------------------------------

\subsection{Results on the 4FGL sample}
We analyzed 31 sources out of those listed in the 4FGL classified as one AGN, 20 BCUs, six BL Lacs, and four FSRQs. 
The AGN appear to be optically classified as a BZQ, while we classified the BCUs as four BZQs and 16 BZBs. For one BZB, we obtained a firm $z$ estimate. Then we confirm all FSRQs as BZQs and their literature redshift measurements, assuming an uncertainty of $\delta z=$0.01 for the literature estimates. We classified four out of six BL Lacs as 3 BZBs confirmed at unknown redshift, and three BZB confirmed at the same $z$ listed in the 4FGL.

Then two known BL Lacs, namely: 4FGL J1410.3+1438 and 4FGL J1503.5+4759, showed a classical quasar-like spectrum leading us to label them as BZQ. 
Also, for one BZQ, 4FGL J1145.5+4423, we found a BZG spectrum in our analysis.
These could be two transitional sources (i.e., changing look blazars), but we cannot confirm their variable nature since we could not retrieve from the literature the optical spectrum used to classify them in the 4FGL.

\subsection{Results on targets pointed during the optical spectroscopic campaign of UGSs and BCUs}

In the sample of sources observed during our optical spectroscopic campaign, we analyzed eight optical sources distinguished as one BZQ, two BZGs, and five BZBs. The summary of our results is reported below.

We classified the BZQ \citep{quest} J010216.63+094411.1 as BZB, and confirmed its redshift. For all BZBs we confirm their class and redshift, four with no redshift, and one at the same $z$ previously known. 
The BZG J163738.24+300506.4 was originally identified as lying at $z=$0.079, however, we report that there is another source within 3\farcs3 observed in LAMOST and SDSS and we classified it as a BZQ at $z=$0.819. 
%I should keep it as BZB  ---< discuss

The other BZG, namely J134243.61+050432.1, is indeed a radio galaxy, a.k.a. 4C +05.57, lying at $z=$0.136 and with extended radio emission at 1.4 GHz.

\subsection{Blazars in the Roma-BZCAT}
The majority of the spectra we investigated are those related to the Roma-BZCAT for a total of 353 blazars. These were all previously classified as 184 BZQs, 39 BZGs, and 130 BZBs. The results of this part of our investigation are described in the following.

A large fraction of BZQs were confirmed in both classification and redshift estimates (i.e., assuming an uncertainty of 0.01 due to the heterogeneous uncertainties found in the literature), 174 sources out of 184 analyzed. For an additional three of them, namely: 5BZQ J1728+3838, 5BZQ J0010+1724, and 5BZQ J1047+2635, we provided a new redshift estimate, different from the previous one. Unfortunately, we could not find a literature spectrum to verify the $z$ estimate previously reported in the Roma-BZCAT to carry out a comparison. 

We confirmed the BZG nature of 23 out of 39 BZGs analyzed, all at the same redshifts previously known. Then we found that 5BZG J0048+3157 (a.k.a. NGC 0262) shows a classical Seyfert 2-like spectrum. 

Five more BZGs, namely: 5BZG J0751+1730, 5BZG J0756+3834, 5BZG J1504-0248, 5BZG J1512+0203, and 5BZG J2346+4024, showed a quasar-like spectrum, and we classified them as BZQs, all at the same redshifts previously known with the only exception of 5BZG J2346+4024 lying at $z=$0.459 instead of $z=$0.0838 as reported in the Roma-BZCAT. These five transitional (i.e., changing look) sources were found in addition to 12 BZGs for which our analysis classifies them as BZBs, one with a different redshift estimate, 5BZG J0916+5238 having a LAMOST featureless spectrum that did not allow us to confirm its previous $z=$0.190 estimate.

Finally, for 130 BZBs, our results are summarized as follows.
\begin{itemize}
\item Two BZBs of them appear to be transitional sources being classified as BZQs 5BZB J1402+1559 and 5BZB J1001+2911 at the same redshift. %For the remaining one, 5BZB J1216-0243, we obtained a different $z$ at 1.01 instead of 0.359.
\item Seventy-nine BZBs with no redshift estimate were all confirmed, but for six cases, we also obtained a $z$ estimates or lower limits, namely for 5BZB 0124+3249 at $z=$0.780, 5BZB 0127-0151 at $z=$0.337, and 5BZB J0910+3902 at $z=$0.199. In the case of 5BZB J1117+5355, 5BZB J1552+0850 and 5BZB J1643+3221, the presence of intervening systems with optical features superimposed on their spectral continuum allowed us to get the following lower limits: 0.721, 1.0160 and 1.029, respectively.
\item Six BZBs are labeled as BL Lac candidates in the Roma-BZCAT, but we were able to obtain a firm BZB classification and, in particular, for two of them: 5BZB J0124+3249 and 5BZB J0127-0151 also a $z$ estimate at 0.780 and 0.337, respectively.
\item We also analyzed optical spectra for 51 BZBs of the Roma-BZCAT with a known redshift estimate, 12 flagged as uncertain. For all these 51 BZBs, we confirmed their BL Lac nature, and two of those with uncertain redshift we could verify their $z$ values: 5BZB J1211+2242 at $z=$0.453, 5BZB J1410+2820 at $z=$0.521, and for 5BZB J0911+2949, we found that it lies at $z=$0.438 instead of $z=$0.446.
\end{itemize}

%------------------------------------------------------------------------------------------------------------------------
%
\begin{table*}
%\resizebox{\textwidth}{!}{
\caption{Summary of results}
\begin{tabular}{lllllll}
\noalign{\smallskip}
\hline
\noalign{\smallskip}
\hline
Orig. Class & Number & BZB(z) & BZB & BZQ & BZG & Other\\
            &        &        &     &     &     &      \\
(1)         & (2)    &  (3)   & (4) & (5) & (6) & (7)  \\
\hline
\multicolumn{5}{c}{4FGL} \\ 
\hline 
AGN         & 1      & \dots  &\dots& 1   &\dots& \dots\\ 
BCU         & 20     & 1      & 15  & 4   &\dots&\dots \\ 
BLL         & 6      & 1      & 3   & 1   & 1   &\dots \\ 
FSRQ        & 4      & \dots  &\dots& 4   &\dots&\dots  \\ 
\hline %checked
\multicolumn{5}{c}{Optical Campaign} \\ 
\hline 
BZQ         & 1      & 1      &\dots&\dots&\dots&\dots  \\ 
BZB         & 5      & 1      & 4   &\dots&\dots&\dots  \\ 
BZG         & 2      & \dots  & \dots & 1&\dots& 1 \\ 
\hline 
\multicolumn{5}{c}{Roma-BZCAT} \\ 
\hline 
BZQ         & 184    & \dots  & 6   & 178 &\dots&\dots  \\ 
BZB         & 130    & 26     & 102 & 2   &\dots&\dots  \\ 
BZG         & 39     & 9      & 1   & 5   & 23  & 1     \\ 
\hline 
\noalign{\smallskip}
\end{tabular}
%}
\label{tab:summary}

Note: columns 
(1) class in the original sample (i.e. 4FGL, Roma-BZCAT or Optical Campaign); 
(2) total number of sources within the original class;
(3) BZB with z measurements from the present analysis; 
(4) BZB without z measurements from the present analysis; 
(5) number of BZQs; 
(6) number of BZGs; 
(7) other type of AGN;
\end{table*}
%------------------------------------------------------------------------------------------------------------------------

\subsection{Changing-look blazars}

Given that the optical classification (BZBs, BZQs, or BZGs) is defined by an arbitrary limits on the equivalent width and continuum to spectral features contrast, their spectral variations \citep{gaur12,leon13,isler13} can lead to changes in their optical classification, as reported in the literature \citep{vermeulen95,pian99,capetti10,ghisellini11,ruan14,optcmpv,acosta17}. We refer as changing-look blazars those objects with different optical spectral classifications reported in two different epochs in contrast with the definition of BL Lac adopted in other works \citep[e.g.,][]{shaw12}. Listing changing-look blazars is important for follow-up observational programs aiming to study their spectral variation.

\begin{figure*}
\begin{center}
\includegraphics[width=\textwidth]{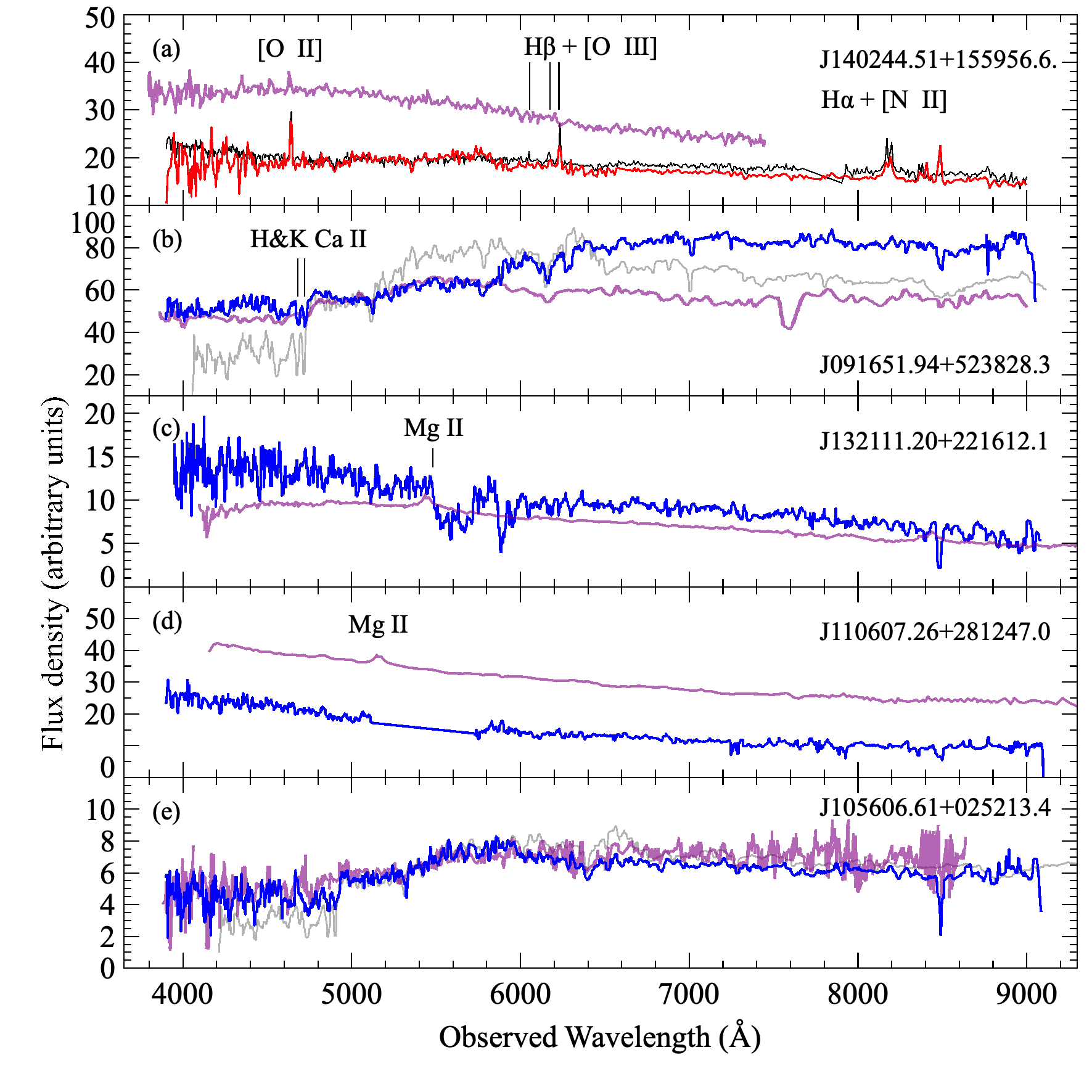}
\end{center}
\caption{
Comparison of the literature spectra (light violet) with the LAMOST spectra (blue if BZB red if BZQ) for those blazars with different classification with available spectral image. 
{\it (a)} the optical spectrum of the changing-look blazar 5BZB J1402+1559, in red the LAMOST spectrum, and in black the SDSS spectrum. We classified this source as BZQ and compared with the BZB spectrum reported by \citet{baldwin77}.
{\it (b)} The spectrum of  5BZG J0916+5238 classified as BZG in the Roma-BzCAT, based on the spectrum of \citet{nass96} compared with the BZB LAMOST spectrum. In gray, we show the spectrum of the elliptical galaxy  J084640.34+281829.6 to compare the Ca {\sc II} break with the blazar one.
{\it (c)} The spectrum of 5BZQ J1321+2216, a BZQ as classified by \citet{shaw12} in comparison to the LAMOST spectrum classified here as BZB.
{\it (d)} The spectrum of 5BZQ J1106+2812, classified as BZQ by \citet{shaw12} compared to the LAMOST spectrum showing a BZB.
{\it (e)} The spectrum of the 5BZG J1056+0252 reported by \citet{fischer98} in comparison to the BZB LAMOST spectrum. In gray, we show the spectrum of the elliptical galaxy J084640.34+281829.6 to compare the Ca {\sc II} break with the blazar one.
}
\label{fig:clb}
\end{figure*}

We compared the literature spectral classification with our classification based on LAMOST spectra. 
For 26 objects, we found a difference in classification. We summarize these results in Table~\ref{tab:clb}. 
We performed a search in the literature for these potential changing-look blazars looking for their spectra and details on their classification. We found the spectra only for five of them, and we compare their literature spectra with the LAMOST one, see Figure~\ref{fig:clb}. 
The five potential changing-look blazars for which we found spectra are:
\begin{itemize}
\item J140244.51+155956.6, originally classified BZB \citep{baldwin77}, in Figure~\ref{fig:clb}, we show the literature spectrum \citet{baldwin77}, in comparison with that collected from LAMOST (2013 February11) and SDSS (2007 June 21). Those spectra available in both SDSS and LAMOST shows intense [O$_ {\,II}$], H$\beta$ and the [O$_ {\,III}$] doublet emission lines letting us classify it as a BZQ.%2007-06-21
\item 5BZG J0916+5238 is a BZG in the Roma-BzCAT, based on the spectrum of \citet{nass96}. We re-classified it as a BZB according to our LAMOST investigation.
\item 5BZQ J1321+2216 was originally classified as a BZQ by \citet{shaw12}. The LAMOST spectrum, although with a lower signal-to-noise ratio, indicates a BZB source type.  However, this difference in classification can be due to observational effects like signal-to-noise ratio.
\item The spectrum of 5BZQ J1106+2812, classified as BZQ by \citet{shaw12}, is indeed BZB according to the LAMOST spectrum. However, we can not compare it with the literature spectrum due to the gap in the LAMOST between $5200-5800$ \AA. Considering this observational effect, we report it as a potential changing-look blazar.
\item 5BZG J1056+0252 was reported by \citet{fischer98} and classified as a BZG in the Roma-BzCAT. We classified it as a BZB according to the LAMOST spectrum. 

\end{itemize}

%However, we report them as changing-look blazar as the literature classification differs from ours.
%Image for these:
%J105606.61+025213.4_1998AN....319..347F
%\item J150407.51-024816.5 classified as BZG in the Roma-BzCAT and previously classified as BZQ by  \citep{sabbey01}.
%J124355.77+404358.4_ 2013AJ....146...10T
%\item 5BZQ J1243+4043, a BZB with a foreground galaxy at $z=0.0172$.
%5BZB J1216-0243 we identified Mg II as the emission line at 5594 \AA\ measuring a redshift $z=1.00$
%J151215.74+020316.9_1991ApJS...75..297H bzcat bzg
%\item 4FGL J1503.5+4759 
%J132111.20+221612.1_2012ApJ...748...49S
%\end{itemize}

The optical classification of blazars can be affected by several observational effects, including: the spectral signal-to-noise ratio, as weak spectral features can be swamp by the noise, spectral coverage, and spectral resolution, to name a few. 
However, despite these sources of uncertainties, on the basis of our analysis we are confident about claiming the changing-look nature of J140244.51+155956.6, J132111.20+221612.1, and J105606.61+025213.4 being less confident for the remaining ones for the reasons previously described.
%We provide a potential classification as changing-look for all the others as their LAMOST spectra are heavily affected by systematic errors, or we could not find the image of the spectrum used for their classification in the literature.

Possible scenarios for accounting for the BZB-BZQ transition objects include that these blazars are:
\begin{itemize}
\item broad-lined objects with highly beamed jets and with radiatively efficient accretion \citep{ruan14,giommi12},
\item quasars with weak radiative cooling and with their broad lines overwhelmed by the non-thermal continuum \citep{ghisellini12}, 
\item  broad-lined objects with variations of their jet bulk Lorentz factor \citep{bianchin09}.
\end{itemize}
%bzb-bzg land10 giommi...
We also considered the transitional objects BZB-BZG as changing-look blazars. BZGs and BZBs are differentiated by the Ca {\sc II} break strength, with an arbitrary limit of $<0.25$ for BZBs. Thus changes in these classes are sensitive to the continuum arising from the jet and the orientation angle \citep{massaro14a,landt02}. 
%In this context, transitional sources, often classified as BL Lacs due to their weak emission lines are not sources with intrinsic low ionization but broad-line objects for which their AGN-thermal emission is highly diluted by the beamed non-thermal emission.

%Additionally, noise in the spectra and the observed spectral range can affect both classifications, based on the equivalent width and the Ca II break.
%giommi 12 blazars are classified into flat-spectrum radio quasars (FSRQs), BL Lacertae (BL Lac) objects, low-synchrotron, or high-synchrotron peaked objects according to a varying mix of the Doppler-boosted radiation from the jet, the emission from the accretion disc, the broad-line region, and the light from the host galaxy.  different classes: intrinsically weak lined objects, more common in X-ray-selected samples, and heavily diluted broad-lined sources, more frequent in radio-selected samples
%The classification of blazars is affected by several selection effects. 
%For example, samples selected in X-rays and $\gamma$-rays tend to have synchrotron peaks at higher frequencies than the synchrotron peak of blazars selected at radio frequencies, independently of their selection method \citep{giommi12}. Another selection effect is the lack of redshift measurements of blazars with high-synchrotron peak.

The optical classification of blazars can also be affected by selection effects.
For example, as suggested by \citet{giommi12}, intrinsically weak-lined blazars are more common in X-ray selected samples (since they are weaker in radio frequencies), while blazars with broad lines are more common in radio selected samples. The first ones are always classified as BZBs (therefore more common in X-ray limited selections), while the second ones will be classified as BZBs only if the continuum is high enough to sink the lines (therefore blazars selected in radio are more likely to change classification, since this essentially depends only on the non-thermal continuum level).  Another selection effect is the lack of redshift measurements of blazars with high-synchrotron peak.
Then broad-line objects are often included in the BL Lac class if their non-thermal continuum swamps the broad components \citep{giommi12,giommi13}.
An example of this variability is shown for the source 5BZB J1402+1559 that changed from BZB to BZQ showing a decrease of the continuum, that lets the line equivalent width increase when comparing the two epochs spectra.
To avoid selection effects, in the literature, there are proposed classifications schemes based on a more physical basis, classifying blazars as low-ionization and high-ionization \citep{giommi12,giommi13,giommi15} or equivalently by a limit on the BLR luminosity $L_{BLR}\sim 5 \times 10^{-4}$ in Eddington units, with $L_{BLR}$ the luminosity of the broad line region \citep{ghisellini11}. 
%Another classification, was proposed by \citet{landt04} to use the [OIII]$\lambda5007-$[OII]$\lambda3727$ equivalent width plane to separate weak- from strong-lined sources. From this plane its is possible to differentiate the intrinsic spectral variability of radio-loud AGNs from the orientation effects. The discrimination is done given that the equivalent width of both lines are dependent on the orientation angle. For sources with viewed at larger angles having lower equivalent widths. 

%------------------------------------------------------------------------------------------------------------------------
\begin{table*}
\caption{\large Changing-look blazars}
%\tiny
%\resizebox{\textwidth}{!}{
\begin{tabular}{llcllclc}
\noalign{\smallskip}
\hline
\noalign{\smallskip}
\hline
  \multicolumn{1}{c}{Name} &
  \multicolumn{1}{c}{z$_1$} &
  \multicolumn{1}{c}{Cat. Class} &
  \multicolumn{1}{c}{Class Reference} &
  \multicolumn{1}{c}{z$_2$}  &
  \multicolumn{1}{c}{Our Class} \\

  \multicolumn{1}{c}{(1)} &
  \multicolumn{1}{c}{(2)} &
  \multicolumn{1}{c}{(3)} &
  \multicolumn{1}{c}{(4)} &
  \multicolumn{1}{c}{(5)}  \\
\hline
\multicolumn{6}{c}{4FGL}  \\
\hline
%  4FGL J1145.5+4423   & 0.2998 & fsrq & \citet{} & 0.300 & bzq\\
  4FGL J1410.3+1438   & 0.1443 & bll  & \citet{4fgl} & 0.144 & bzq\\%in sdss
  4FGL J1503.5+4759   & 0.3445 & bll  & \citet{4fgl} & 0.344 & bzq\\
\hline
\multicolumn{6}{c}{Optical Campaign}  \\
\hline
  J134240.02+094752.4 & 0.2828 & bzq  & \citet{optcmpvi} & 0.283 & bzb\\
%  J163738.24+300506.4 & 0.0786 & bzg  & \citet{pena20} & 0.819 & bzq\\
\hline
\multicolumn{6}{c}{Roma BzCAT}  \\
\hline
  5BZG J0006+1051     & 0.168  & bzg  & \citet{bzcat5} & 0.168 & bzb\\
  5BZG J0022+0006     & 0.306  & bzg  & \citet{bzcat5} & 0.306 & bzb\\
%  5BZG J0110+4149     & 0.096  & bzg  & \citet{} & 0.096 & bzg\\
  5BZG J0303+0554     & 0.196  & bzg  & \citet{bzcat5} & \dots & bzb\\
  5BZG J0751+1730     & 0.187  & bzg  & \citet{bzcat5} & 0.186 & bzq\\
  5BZG J0756+3834     & 0.216  & bzg  & \citet{bzcat5} & 0.216 & bzq\\
  5BZG J0916+5238     & 0.19   & bzg  & \citet{nass96} & 0.190 & bzb\\
  5BZB J1001+2911     & 0.558  & bzb  & \citet{bzcat5} & 0.556 & bzq\\
  5BZQ J1043+2408     & 0.56   & bzq  & \citet{bzcat5} & \dots & bzb\\
  5BZQ J1054+3855     & 1.363  & bzq  & \citet{bzcat5} & \dots & bzb\\
  5BZG J1056+0252     & 0.236  & bzg  & \citet{fischer98} & 0.239 & bzb\\
  5BZG J1103+0022     & 0.275  & bzg  & \citet{bzcat5} & 0.275 & bzb\\
  5BZQ J1106+2812     & 0.844  & bzq  & \citet{shaw12} & \dots & bzb\\
%  5BZB J1216-0243     & 0.359  & bzb  & \citet{} & 1.005 & bzq\\
%  5BZG J1216+0929     & 0.0935 & bzg  & \citet{} & 0.094 & bzb\\
  5BZQ J1243+4043     & 1.518  & bzq  & \citet{bzcat5} & \dots & bzb\\%fg gal at 0.0172
  5BZQ J1321+2216     & 0.943  & bzq  & \citet{bzcat5} & \dots & bzb\\
  5BZG J1326+1229     & 0.204  & bzg  & \citet{bzcat5} & 0.206 & bzb\\
  5BZQ J1343+2844     & 0.905  & bzq  & \citet{evans04} & \dots & bzb\\
  5BZB J1402+1559     & 0.244  & bzb  & \citet{baldwin77} & 0.245 & bzq\\
  5BZG J1449+2746     & 0.227  & bzg  & \citet{bzcat5} & 0.227 & bzb\\
  5BZG J1504-0248     & 0.217  & bzg  & \citet{sabbey01} & 0.217 & bzq\\
  5BZG J1512+0203     & 0.22   & bzg  & \citet{hewitt91} & 0.220 & bzq\\
  5BZG J1730+3714     & 0.204  & bzg  & \citet{bzcat5} & 0.205 & bzb\\
  5BZG J1733+4519     & 0.317  & bzg  & \citet{bzcat5} & 0.317 & bzb\\
  5BZG J2346+4024     & 0.0838 & bzg  & \citet{sowards05} & 0.459 & bzq\\
\hline 
\hline 
\noalign{\smallskip} 
\end{tabular}%} 
\label{tab:clb}\par 
Note: Columns (1) Source name ; (2) redshift in the sample catalog; (3) redshift reference; and,(4)redshift identified thanks to our analysis. 
\end{table*}
%------------------------------------------------------------------------------------------------------------------------

\subsection{New Redshift measurements}

We measured the redshift for 15 blazars, ten without previous measurements reported in the literature and five for which the literature redshift do not match with our measurements. We summarize these measurements in Table~\ref{tab:newz}. For three of the new measurements we could estimate lower limits, namely 5BZB J1117+5355, 5BZB J1552+0850, and 5BZB J1643+3221 at z$>0.721$, z$>1.016$, and z$>1.029$. 
We did not discover high redshift BZBs. All redshift estimates obtained are consistent with the known BZB's redshift distribution.

%J001033.99+172418.7 reported at $z=1.601$ by \citep{wills76} based on the observed emission line at 4961 \AA\. We re-identified this line as Mg II from the LAMOST spectrum.  We additionally identified H$\beta$ and [O III] measuring the redshift at $z=0.7691$.

%------------------------------------------------------------------------------------------------------------------------
\begin{table*}
\caption{New or updated redshift measurements}
%\tiny
%\resizebox{\textwidth}{!}{
\begin{tabular}{llcllclc}
\noalign{\smallskip}
\hline
\noalign{\smallskip}
\hline
  \multicolumn{1}{c}{Name} &
  \multicolumn{1}{c}{z$_1$} &
  \multicolumn{1}{c}{Reference} &
  \multicolumn{1}{c}{z$_2$}  \\

  \multicolumn{1}{c}{(1)} &
  \multicolumn{1}{c}{(2)} &
  \multicolumn{1}{c}{(3)} &
  \multicolumn{1}{c}{(4)}  \\
\hline
\multicolumn{4}{c}{4FGL}  \\
\hline
4FGL J0135.1+0255        &  \dots   & \dots            &  0.372\\
4FGL J0156.5+3914        &  \dots   & \dots            &  0.446\\
%4FGL J0904.0+2724        &  1.7237  & \citet{hewett10} &  1.716\\
4FGL J1418.4+3543        &  \dots   & \dots            &  0.819\\
4FGL J2207.6+0053        &  \dots   & \dots            &  0.970\\
\hline
\multicolumn{4}{c}{Optical Campaign}  \\
\hline
J163738.24+300506.4      &  \dots   & \citet{pena20}            &  0.819\\
\hline
\multicolumn{4}{c}{Roma BzCAT}  \\
\hline
5BZQ J0010+1724          &  1.601   & \citet{wills76}  &  0.769\\
%5BZQ J0108+0135          &  2.099   & \citet{hewett95} &  2.109\\ 
5BZB J0124+3249          &  \dots   & \dots            &  0.780\\
%5BZQ J0124+2805          &  0.71    & \citet{hook96}   &  0.796\\
%5BZQ J0125+0146          &  1.559   & \citet{perlman98}&  1.570\\
5BZB J0127-0151          &  \dots   & \dots            &  0.337\\
%5BZQ J0148+3854          &  1.442   &\citet{henstock97}&  1.427\\
%5BZQ J0821+5031          &  2.1326  & \citet{hewett10} &  2.122\\
5BZB J0910+3902          &  \dots   & \dots            &  0.199\\
%5BZQ J0930+4644          &  2.0360  & \citet{hewett10} &  2.015\\
5BZQ J1047+2635          &  0.99    & \citet{healey08} &  2.560\\
%5BZG J1053+4929          &  0.14    & \citet{sdssdr3}  &  0.14\\
5BZB J1117+5355          &  \dots   & \dots            &  $>0.721$\\
%5BZB J1216-0243          &  0.359   & \citet{bauer00}  &  1.005\\
%5BZQ J1319+4851          &  1.17    & \citet{sowards05}&  1.156\\
%5BZQ J1459+4442          &  3.401   & \citet{sowards05}&  3.428\\
5BZB J1552+0850          &  \dots   & \dots            &  $>1.016$\\
5BZB J1643+3221          &  \dots   &             &  $>1.029$\\
5BZQ J1728+3838          &  1.386   & \citet{monroe16} &  0.630\\
%5BZQ J2330+1100          &  1.489   & \citet{healey08} &  1.501\\
5BZG J2346+4024          &  0.0838  & \citet{sowards05}&  0.459\\
%there are sources with redshift in the literature but not reported in the sample catalog 
\hline
\hline
\noalign{\smallskip}
\end{tabular}%}
\label{tab:newz}\par
Note: Columns (1) Source name ; (2) redshift in the sample catalog; (3) redshift reference; and,(4)redshift identified thanks to our analysis.
\end{table*}
%------------------------------------------------------------------------------------------------------------------------

\section{Summary and conclusions}
\label{sec:summary}
Blazars are the rarest class of active galactic nuclei and among them the subclass of BL Lacs is certainly the most elusive one \citep{urry95} even in the $\gamma$-ray sky where they constitute the largest known population of associated sources \citep{4fgl}.

Here we present the results of the analysis of 392 blazar spectra available in the archive of LAMOST, using its data release 5. Sources were selected out of three main samples or catalogs, namely (1) the Fourth \fer-LAT Point Source Catalog \citep{4fgl}, (2) the list of targets recently pointed during the follow up spectroscopic campaign of unidentified/unassociated $\gamma$-ray sources \citep{massaro13c,dabrusco14,optcmpiii,optcmpiv,optcmpv,optcmpvi,optcmpvii,optcmpviii,optcmpix,optcmpx,pena20}, including results of other groups \citep[e.g.,][]{shaw13,klindt17,paiano17b,paiano17c,paiano17d} and (3) the multifrequency catalog of blazars: the Roma-BZCAT \citep{bzcat1,bzcat5}. The total number of those lying in the LAMOST footprint and having an optical spectrum with a signal to noise ratio greater than 10, thus allowing us to perform our investigation, are a total of 392. For $\sim$48\%, i.e., 188 out of 392, we also compare the spectra available thanks to LAMOST with those collected in the SDSS. 

%for 102 objects agree the redshift automatic measurement between LAMOST and SDSS
From the 4FGL sample, we analyzed the spectra of 31 sources, including 20 BCUs, ten known blazars, and one AGN. Our largest sample, analyzed here, is constituted by 353 blazars in Roma-BZCAT, classified as 184 BZQs, 39 BZGs, and 130 BZBs. Our results are then summarized in Table~\ref{tab:summary}, comparing the original classification with the classification presented in the following analysis.
\begin{itemize}
    \item We classified 20 BCUs listed in the 4FGL and one AGN, all of unknown nature,  confirming that they are all blazars. Given the large number (15 out of 19) BCUs that are indeed BZBs, we conclude that BL Lacs are the most elusive class of extragalactic $\gamma$-ray sources.
    \item We obtained a total of 15 new redshifts for blazars and previously uncertain blazars listed in our samples.
    \item We discovered 26 potential transitional (i.e., changing-look) blazars. In particular we found both BZBs and BZGs that were reclassified as BZQs as well as BZQs that showed a classical BZB featureless optical spectrum. Although, we did not find literature spectra to firmly classify them as changing-look.
    \item We were also able to confirm the blazar-like nature of six unconfirmed BL Lac listed in the 4FGL.
\end{itemize}
All the remaining sources analyzed are in agreement with previous literature classifications. In particular the redshifts for six BZBs, all from the Roma-BZCAT sample, are listed here: 
5BZB J0124+3249 at $z=$0.780,
5BZB J0127-0151 at $z=$0.337,
5BZB J0910+3902 at $z=$0.199,
5BZB J1117+5355 at $z=$0.721,
5BZB J1552+0850 at $z=$1.016, and,
5BZB J1643+3221 at $z=$1.029, 
since they all lie in the tail of high redshift BL Lac population \citep{landoni18} but consistent with the whole distribution as in Roma-BZCAT and similar to those arising from our optical spectroscopic campaign \citep{pena20}.

From the comparison of the literature spectral classification with our classification based on LAMOST spectra we found a difference in classification for 26 objects. We summarize these results in Table~\ref{tab:clb}, showing their original classification and its reference. 
%deatils of this result 
Among these 26 blazars we confirm the changing-look nature of J140244.51+155956.6. We present the comparison of the literature spectrum with the LAMOST and SDSS spectra. In Figure~\ref{fig:clb}, we also present other potential changing-look blazars for which we found literature spectra.
Studying the population of changing-look blazars is of crucial importance for avoiding selection effects due to miss classify objects for studying the cosmic evolution of blazars. This sample of changing-look blazars can be of interest of spectral variability studies as the Time Domain Spectroscopic Survey \citep{green14,macLeod18}.

LAMOST survey revealed to be an extremely useful tool, previously unexplored, to investigate blazars, their variability, classification and redshift measurement.
%Since during the preparation of the present manuscript LAMOST DR6 has been publicly released to the astronomical community, we aim at exploring it in a forthcoming paper also investigating the possibility to discover new BL Lac objects among the unidentified $\gamma$-ray sources adopting the same procedure we previously developed and successfully implemented \citep{}.

\vspace{10mm}

\begin{acknowledgements}
%We thank the referee for useful comments that led to improvements in the paper.
% friends and colleagues
H.P.-H. and V.C. acknowledge support from CONACyT research grant No. 280789.
MFG acknowledges support from the National Science Foundation of China (grant 11873073).
% grants
%This investigation is supported by 
This work is supported by the ``Departments of Excellence 2018 - 2022’’ Grant awarded by the Italian Ministry of Education, University and Research (MIUR) (L. 232/2016). This research has made use of resources provided by the Ministry of Education, Universities and Research for the grant MASF\_FFABR\_17\_01. A.P. acknowledges financial support from the Consorzio Interuniversitario per la fisica Spaziale (CIFS) under the agreement related to the grant MASF\_CONTR\_FIN\_18\_02.

%LAMOST
This work is based on data from Guoshoujing Telescope (the Large Sky Area Multi-Object Fiber Spectroscopic Telescope LAMOST), is a National Major Scientific Project built by the Chinese Academy of Sciences. Funding for the project has been provided by the National Development and Reform Commission. LAMOST is operated and managed by the National Astronomical Observatories, Chinese Academy of Sciences.
% ASDC
Part of this work is based on archival data, software or on-line services provided by the ASI Science Data Center.
% HEASARC
%This research has made use of data obtained from the high-energy Astrophysics Science Archive
%Research Center (HEASARC) provided by NASA's Goddard Space Flight Center; 
% SIMBAD and NED
the SIMBAD database operated at CDS,
Strasbourg, France; the NASA/IPAC Extragalactic Database
(NED) operated by the Jet Propulsion Laboratory, California
Institute of Technology, under contract with the National Aeronautics and Space Administration.
% NVSS
%Part of this work is based on the NVSS (NRAO VLA Sky Survey):
%The National Radio Astronomy Observatory is operated by Associated Universities,
%Inc., under contract with the National Science Foundation and on the VLA low-frequency Sky Survey (VLSS).
% SUMSS
%The Molonglo Observatory site manager, Duncan Campbell-Wilson, and the staff, Jeff Webb, Michael White and John Barry, 
%are responsible for the smooth operation of Molonglo Observatory Synthesis Telescope (MOST) and the day-to-day observing programme of SUMSS. 
%The SUMSS survey is dedicated to Michael Large whose expertise and vision made the project possible. 
%The MOST is operated by the School of Physics with the support of the Australian Research Council and the Science Foundation for Physics within the University of Sydney.
% WISE
This publication makes use of data products from the Wide-field Infrared Survey Explorer, 
which is a joint project of the University of California, Los Angeles, and 
the Jet Propulsion Laboratory/California Institute of Technology, 
funded by the National Aeronautics and Space Administration.
% 2MASS
%This publication makes use of data products from the Two Micron All Sky Survey, which is a joint project of the University of 
%Massachusetts and the Infrared Processing and Analysis Center/California Institute of Technology, funded by the National Aeronautics 
%and Space Administration and the National Science Foundation.
%USNO
%This research has made use of the USNOFS Image and Catalogue Archive
%operated by the United States Naval Observatory, Flagstaff Station
%(http://www.nofs.navy.mil/data/fchpix/).
% SDSS
Funding for the SDSS and SDSS-II has been provided by the Alfred P. Sloan Foundation, 
the Participating Institutions, the National Science Foundation, the U.S. Department of Energy, 
the National Aeronautics and Space Administration, the Japanese Monbukagakusho, 
the Max Planck Society, and the Higher Education Funding Council for England. 
The SDSS Web Site is http://www.sdss.org/.
The SDSS is managed by the Astrophysical Research Consortium for the Participating Institutions. 
The Participating Institutions are the American Museum of Natural History, 
Astrophysical Institute Potsdam, University of Basel, University of Cambridge, 
Case Western Reserve University, University of Chicago, Drexel University, 
Fermilab, the Institute for Advanced Study, the Japan Participation Group, 
Johns Hopkins University, the Joint Institute for Nuclear Astrophysics, 
the Kavli Institute for Particle Astrophysics and Cosmology, the Korean Scientist Group, 
the Chinese Academy of Sciences (LAMOST), Los Alamos National Laboratory, 
the Max-Planck-Institute for Astronomy (MPIA), the Max-Planck-Institute for Astrophysics (MPA), 
New Mexico State University, Ohio State University, University of Pittsburgh, 
University of Portsmouth, Princeton University, the United States Naval Observatory, 
and the University of Washington.
% TOPCAT
TOPCAT\footnote{\underline{http://www.star.bris.ac.uk/$\sim$mbt/topcat/}} 
\citep{taylor05} for the preparation and manipulation of the tabular data and the images.
% ALADIN
%The Aladin Java applet\footnote{\underline{http://aladin.u-strasbg.fr/aladin.gml}}
%was used to create the finding charts reported in this paper \citep{bonnarell00}. It can be started from the CDS (Strasbourg - France), from the CFA (Harvard - USA), from the ADAC (Tokyo - Japan), from the IUCAA (Pune - India), from the UKADC (Cambridge - UK), or from the CADC (Victoria - Canada).
\end{acknowledgements}
\clearpage

%\bibliography{sample63}{}
\bibliographystyle{aasjournal}
\bibliography{optcmp} % your references Yourfile.bib



\end{document}